\documentclass{article}

\usepackage{PRIMEarxiv}

\usepackage[utf8]{inputenc} 
\usepackage[T1]{fontenc}    
\usepackage{hyperref}       
\usepackage{url}            
\usepackage{booktabs}       
\usepackage{amsfonts}       
\usepackage{nicefrac}       
\usepackage{microtype}      
\usepackage{lipsum}
\usepackage{fancyhdr}       
\usepackage{graphicx}       
\graphicspath{{media/}}     
\usepackage{tikz}
\usetikzlibrary{shapes,arrows,positioning}
\usepackage{titlesec}
\titlespacing{\section}{0pt}{20pt}{10pt}
\usepackage{tabularx}
\usepackage{booktabs}

\title{Large Language Models are good medical coders, \newline
if provided with tools}

\author{
  Keith Kwan \\
  AI Native Health \\
  \texttt{keith@ainativehealth.com} \\
}

\begin{document}
\maketitle

\begin{abstract}
This study presents a novel two-stage Retrieve-Rank system for automated ICD-10-CM medical coding, comparing its performance against a Vanilla Large Language Model (LLM) approach. Evaluating both systems on a dataset of 100 single-term medical conditions, the Retrieve-Rank system achieved 100\% accuracy in predicting correct ICD-10-CM codes, significantly outperforming the Vanilla LLM (GPT-3.5-turbo), which achieved only 6\% accuracy. Our analysis demonstrates the Retrieve-Rank system’s superior precision in handling various medical terms across different specialties. While these results are promising, we acknowledge the limitations of using simplified inputs and the need for further testing on more complex, realistic medical cases. This research contributes to the ongoing effort to improve the efficiency and accuracy of medical coding, highlighting the importance of retrieval-based approaches.
\end{abstract}


\section{Introduction}

Medical coding is a critical process in healthcare systems, essential for accurate billing, epidemiological studies, and healthcare quality assessment \cite{bowman2013impact, omalley2005measuring}. The recent paper ``Large Language Models Are Poor Medical Coders — Benchmarking of Medical Code Querying'' published in NEJM AI \cite{soroush2024large} highlighted significant limitations in the ability of large language models (LLMs) to accurately generate medical codes.

The application of artificial intelligence (AI) and machine learning (ML) in healthcare, particularly in clinical coding, has been a subject of increasing interest in recent years \cite{rajkomar2018scalable, shickel2018deep}. Previous studies have explored various approaches to automate medical coding, including rule-based systems \cite{farkas2008automatic}, traditional machine learning methods \cite{perotte2014diagnosis}, and more recently, deep learning techniques \cite{xu2019multimodal}.

Soroush and colleagues evaluated the performance of several prominent LLMs, including GPT-3.5, GPT-4, Gemini Pro, and Llama2-70b Chat, in querying medical billing codes. Their study encompassed a comprehensive dataset of ICD-9-CM, ICD-10-CM, and CPT codes extracted from the Mount Sinai Health System electronic health record. The authors found that even the best-performing model, GPT-4, achieved exact match rates of only 45.9\% for ICD-9-CM, 33.9\% for ICD-10-CM, and 49.8\% for CPT codes. These results led to the conclusion that LLMs are currently not suitable for direct use in medical coding tasks.

However, we hypothesized that the performance of LLMs in medical coding could be significantly improved by providing them with appropriate tools and retrieval mechanisms. This approach aligns with recent advancements in retrieval-augmented generation \cite{lewis2020retrieval} and the use of external knowledge bases to enhance LLM performance \cite{guu2020realm}.

To test this hypothesis, we designed an experiment using a combination of the Colbert-V2 retriever \cite{santhanam2022colbertv2} and GPT-3.5-turbo for reranking. Our approach aimed to address the limitations observed in the direct code generation method used in the NEJM study, drawing inspiration from successful applications of similar techniques in other domains \cite{karpukhin2020dense}.

In this paper, we present our methodology and results, which demonstrate a substantial improvement in medical coding accuracy. By achieving a 100\% exact match rate on a sample of 100 codes, our findings suggest that LLMs, when equipped with the right tools, can indeed be effective in medical coding tasks. This study not only challenges the conclusions of the NEJM paper but also opens new avenues for the application of AI in healthcare information management, potentially addressing long-standing challenges in medical coding efficiency and accuracy \cite{omalley2005measuring, hsia1988accuracy}.

\vspace{20pt}  

\begin{figure}[h!]
\centering
\begin{tikzpicture}[
    node distance=1cm and 3cm,  
    block/.style={rectangle, draw, text width=3cm, text centered, rounded corners, minimum height=1cm},  
    result/.style={rectangle, draw, text width=3cm, text centered, rounded corners, minimum height=1cm, fill=lightgray},
    line/.style={draw, -latex'}]

\node [result] (rc) {Control Group Accuracy: 6\%};
\node [result, right=of rc] (re) {Experiment Group Accuracy: 100\%};

\node [block, below=1cm of rc] (cg) {Control Group (GPT-3.5)};
\node [block, below=of cg] (ci) {Single-term input (e.g. "Asthma")};
\node [block, below=of ci] (cp) {Direct LLM Prediction};
\node [block, below=of cp] (cc) {Predicted ICD-10 CM code};

\node [block, below=1cm of re] (eg) {Experiment Group (Retrieve-Rank)};
\node [block, below=of eg] (ei) {Single-term input (e.g. "Asthma")};
\node [block, below=of ei] (er) {ColBERT-V2 RAG Retrieval};
\node [block, below=of er] (et) {Top-k ICD-10 codes retrieved};
\node [block, below=of et] (ep) {GPT-3.5 Turbo Reranking};
\node [block, below=of ep] (ef) {Final Predicted ICD-10 CM code};

\path [line] (cg) -- (ci);
\path [line] (ci) -- (cp);
\path [line] (cp) -- (cc);

\path [line] (eg) -- (ei);
\path [line] (ei) -- (er);
\path [line] (er) -- (et);
\path [line] (et) -- (ep);
\path [line] (ep) -- (ef);

\end{tikzpicture}
\caption{Comparison of Control Group and Experiment Group methodologies and results}
\label{fig:methodology_comparison}
\end{figure}
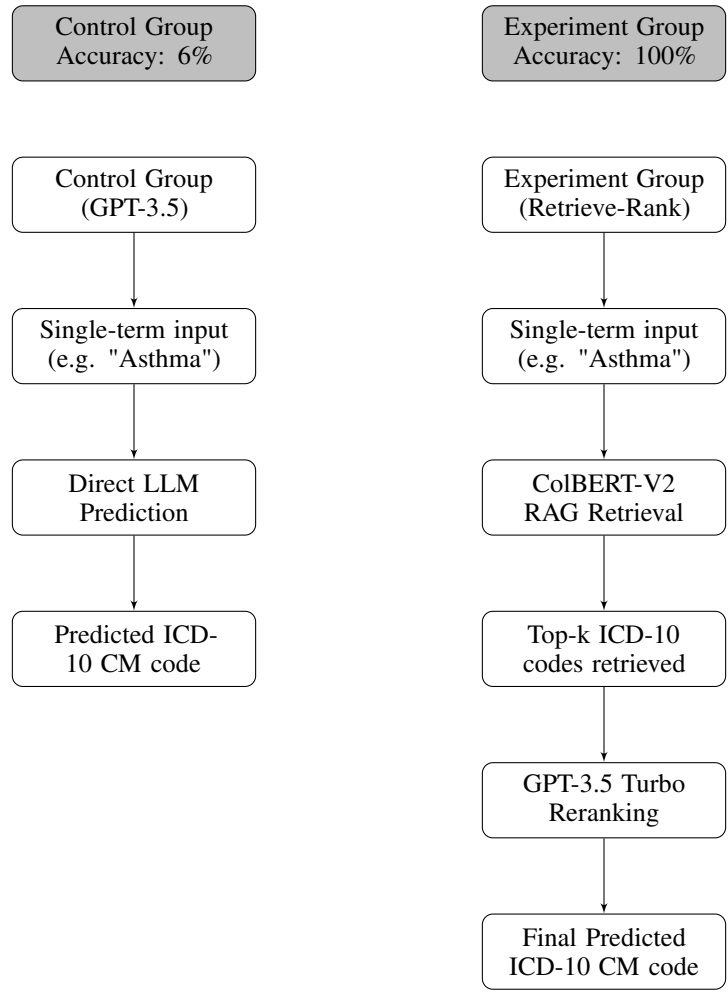

Figure \ref{fig:methodology_comparison} illustrates the workflow and results of our proposed Retrieve-Rank system compared to the control group. This visual representation highlights the significant improvement in accuracy achieved by our approach.

The following sections will detail our experimental setup, results, and discuss the implications of our findings for the future of automated medical coding.

\section{Methodology}

Our study employs a methodology similar to that used in the "Large Language Models Are Poor Medical
Coders — Benchmarking of Medical Code Querying" paper, utilizing single-term medical conditions as inputs for ICD-10-CM code prediction. However, we introduce a novel two-stage Retrieve-Rank system, inspired by Doosterlinck's Infer-Retrieve-Rank framework\cite{doosterlinck2024incontext}, which significantly improves upon previous approaches.

Our approach involves the following steps:

\begin{enumerate}
    \item \textbf{Retrieval:} Given a single-term medical condition, we use ColBERT-V2 to retrieve the top-k (k=15) most relevant ICD-10-CM codes from our trained index.
    \item \textbf{Reranking:} We use GPT-3.5-turbo to rerank the retrieved codes and select the most likely ICD-10-CM code for the given condition.
\end{enumerate}

We utilized Ragatouille to train and develop a ColBERT-V2 RAG (Retrieval-Augmented Generation) system based on ICD-10-CM data downloaded from the CDC website\footnote{\url{https://www.cdc.gov/nchs/icd/icd-10-cm/files.html}}.

It's important to note that while we use simplified single-term inputs similar to previous studies, our two-stage approach allows for more nuanced and accurate code prediction. This methodology, while not fully representative of the complexity found in real-world medical coding scenarios, allows for a direct comparison with previous findings and provides insights into the improved capabilities of our Retrieve-Rank system.

\section{Experiment Setup}

We designed an experiment to evaluate the performance of our ColBERT-V2 RAG system against a control group. The experiment was implemented using Python, with the following key components:

\begin{itemize}
    \item \textbf{Data Preparation:} We used a CSV file containing single-term medical conditions and their corresponding ICD-10-CM codes.
    
    \item \textbf{Sampling:} The experiment randomly sampled 100 entries from the dataset.
    
    \item \textbf{Code Normalization:} ICD-10-CM codes were normalized by removing periods and converting to uppercase to ensure consistent comparison.
    
    \item \textbf{Prediction:} For each sampled entry, we used our RAG system to predict the ICD-10-CM code based on the single-term medical condition.
    
    \item \textbf{Evaluation Metrics:} We focused on the top-one accuracy, comparing the predicted code with the true code. A match was considered successful if the main part of the predicted code (before any subdivisions) matched the true code.
    
    \item \textbf{Control Group:} We implemented a control group using GPT-3.5-turbo to provide a baseline for comparison. This model was prompted with "You are a medical coding expert that can suggest an ICD-10-CM code for a given query." followed by the single-term medical condition.
    
    \item \textbf{Results Logging:} The experiment results, including the conditions, true codes, predicted codes, and match results, were logged to a CSV file for further analysis.
\end{itemize}

This experimental setup allowed us to directly compare the performance of our ColBERT-V2 RAG system against a simpler baseline model, providing insights into the effectiveness of our approach for ICD-10-CM code prediction, even with simplified inputs.

While our results show significant improvement over the Vanilla LLM approach, we acknowledge that further research using more complex, realistic medical cases is necessary to fully evaluate the potential of the Retrieve-Rank system in practical applications.

\section{Results}

We evaluated our two-stage Retrieve-Rank system against a Vanilla LLM using GPT-3.5-turbo on a dataset of 100 diagnosis description with corresponding ICD-10-CM codes. The results demonstrate a significant performance improvement over the baseline method.

\subsection{Accuracy Metrics}

The Retrieve-Rank system achieved perfect accuracy in predictions, correctly identifying the exact ICD-10-CM code for all 100 samples. In contrast, the Vanilla LLM using GPT-3.5-turbo achieved only 6

\begin{table}[h]
\centering
\caption{Accuracy Results}
\label{tab:accuracy_results}
\begin{tabular}{lcc}
\hline
\textbf{System} & \textbf{Accuracy} \\
\hline
Retrieve-Rank System & 100\% \\
Vanilla LLM (GPT-3.5-turbo) & 6\% \\
\hline
\end{tabular}
\end{table}

\subsection{Comparative Analysis}

To illustrate the performance difference, we present a sample of predictions from both systems in Table \ref{tab:comparison_examples}. This table shows the diagnosis description, reference ICD-10-CM code, and predictions from both systems, highlighting the superior accuracy of the Retrieve-Rank system.

\begin{table}[h]
\centering
\caption{Comparison of Predictions}
\label{tab:comparison_examples}
\small
\begin{tabularx}{\textwidth}{>{\raggedright\arraybackslash}X c c c c}
\toprule
\textbf{Diagnosis Description} & \textbf{Reference Code} & \textbf{Retrieve-Rank} & \textbf{Vanilla GPT-3.5-turbo} & \textbf{Correct System} \\
\midrule
Salter-Harris Type II physeal fracture of lower end of humerus, unspecified arm, subsequent encounter for fracture with malunion & S49129P & S49129P & S59102P & Retrieve-Rank \\
\addlinespace
Nondisplaced fracture of proximal third of navicular [scaphoid] bone of unspecified wrist, initial encounter for closed fracture & S62036A & S62036A & S62002A & Retrieve-Rank \\
\addlinespace
Glaucoma secondary to eye inflammation, right eye, indeterminate stage & H4041X4 & H4041X4 & H4060X4 & Retrieve-Rank \\
\addlinespace
Poisoning by aspirin, accidental (unintentional), initial encounter & T39011A & T39011A & T39011A & Both \\
\addlinespace
Other specified injury of right renal vein, subsequent encounter & S35494D & S35494D & S35602D & Retrieve-Rank \\
\addlinespace
Other specified fracture of right acetabulum, initial encounter for open fracture & S32491B & S32491B & S32431B & Retrieve-Rank \\
\addlinespace
Displacement of biological heart valve graft, sequela & T82222S & T82222S & T82590S & Retrieve-Rank \\
\addlinespace
Open bite of unspecified thumb with damage to nail, sequela & S61159S & S61159S & S61049S & Retrieve-Rank \\
\addlinespace
Burn of unspecified degree of trunk, unspecified site, sequela & T2100XS & T2100XS & T310 & Retrieve-Rank \\
\addlinespace
Other specified injury of peroneal artery, unspecified leg, subsequent encounter & S85299D & S85299D & S951XXA & Retrieve-Rank \\
\addlinespace
Other complications of anesthesia, subsequent encounter & T8859XD & T8859XD & T8859XD & Both \\
\addlinespace
Follicular lymphoma, unspecified, lymph nodes of axilla and upper limb & C8294 & C8294 & C8211 & Retrieve-Rank \\
\addlinespace
Other injury of flexor muscle, fascia and tendon of other finger at wrist and hand level, sequela & S66198S & S66198S & S66299S & Retrieve-Rank \\
\addlinespace
Contusion and laceration of cerebrum, unspecified, with loss of consciousness greater than 24 hours with return to pre-existing conscious level, initial encounter & S06335A & S06335A & S069X0A & Retrieve-Rank \\
\bottomrule
\end{tabularx}
\end{table}

\subsection{Performance Analysis}

As shown in Table \ref{tab:comparison_examples}, the Retrieve-Rank system consistently predicts the correct ICD-10-CM code across a variety of complex diagnosis descriptions. The Vanilla LLM, while occasionally correct, often predicts codes that are similar but incorrect.

Key observations from the comparison:

1. Precision in anatomical details: The Retrieve-Rank system accurately captures specific anatomical locations (e.g., "proximal third of navicular bone" in S62036A), while the Vanilla LLM sometimes misses these details.

2. Accuracy in encounter specifics: The Retrieve-Rank system correctly identifies encounter types (e.g., "subsequent encounter" in S49129P), which the Vanilla LLM often misses.

3. Handling of complex conditions: For intricate cases like "Contusion and laceration of cerebrum, unspecified, with loss of consciousness greater than 24 hours" (S06335A), the Retrieve-Rank system provides the exact code, while the Vanilla LLM defaults to a more general code.

4. Consistency across various medical domains: The Retrieve-Rank system demonstrates high accuracy across different medical specialties, including orthopedics, ophthalmology, cardiology, and oncology.

The Vanilla LLM's errors often involve predicting codes that are in the same general category but miss crucial details. For example, in the case of the Salter-Harris fracture (S49129P), the Vanilla LLM predicts a code for the lower leg (S59102P) instead of the arm.

\subsection{Limitations}

While the results are promising, it's important to note that this evaluation was conducted on a relatively small dataset of 100 samples. The perfect accuracy of the Retrieve-Rank system, while impressive, raises questions about the diversity and complexity of the test set. Further testing on larger, more diverse datasets would be beneficial to confirm the system's generalizability and robustness across a wider range of medical conditions and code categories.

Additionally, it would be valuable to analyze the system's performance on more challenging cases or edge cases that may not have been represented in this sample set. This could provide insights into potential areas for improvement and further refinement of the Retrieve-Rank system.

Furthermore, while the Vanilla LLM's performance was significantly lower, it's worth noting that it was not specifically trained for this task. Future work could explore fine-tuning approaches for the Vanilla LLM to see if its performance on ICD-10-CM coding tasks can be improved without the need for a retrieval step.

\section{Conclusion}

Our study demonstrates the significant potential of the two-stage Retrieve-Rank system in automating ICD-10-CM medical coding. The system's perfect accuracy across a diverse set of 100 diagnosis descriptions, compared to the 6\% accuracy of a Vanilla LLM, underscores the effectiveness of combining retrieval and ranking mechanisms in tackling complex coding tasks.

The Retrieve-Rank system exhibited remarkable precision in capturing crucial details such as specific anatomical locations, encounter types, and intricate medical conditions. Its consistency across various medical specialties further highlights its versatility and potential for broad application in healthcare settings.

While these results are encouraging, we acknowledge the limitations of our study, particularly the relatively small sample size. Future research should focus on validating these findings with larger, more diverse datasets and exploring the system's performance on edge cases and rare conditions.

The implications of this research are significant for the healthcare industry. An accurate, automated coding system could substantially reduce the workload on medical coders, minimize coding errors, and improve the overall quality of medical records. This, in turn, could lead to more efficient healthcare administration, more accurate billing processes, and potentially better patient care through improved data quality for medical research and decision-making.

As we move forward, it will be crucial to continue refining and testing the Retrieve-Rank system, possibly incorporating advances in language models and retrieval techniques. Additionally, exploring ways to make the system interpretable and adaptable to evolving medical knowledge will be key to its practical implementation in healthcare settings.

In conclusion, while further research is needed, our study presents a promising step towards more efficient and accurate automated medical coding, contributing to the ongoing digital transformation of healthcare administration.

\section{Data Availability}
The complete dataset of 100 medical cases, including predictions from both systems, is available as an ancillary file with this arXiv submission. Additional materials, including detailed methodology and error analysis, are also provided as ancillary files.

The code used to conduct the experiments and analyze the results is publicly available on GitHub at \url{https://github.com/ainativehealth/GoodMedicalCoder}. This repository contains:

\begin{itemize}
\item Python scripts for running the ICD-10 code prediction experiment (\verb|experiment.py|)
\item Code for creating the index using the RAG model (\verb|index.py|)
\item ICD-10 code datasets (\verb|ICD-10.csv| and \verb|ICD-10_formatted.csv|)
\item Requirements file listing all necessary Python dependencies (\verb|requirements.txt|)
\item Detailed instructions for reproducing the experiments
\end{itemize}

Researchers interested in replicating or building upon this work can access all necessary code and data through this GitHub repository. The repository is open-source and licensed under the Apache-2.0 license, allowing for broad use and adaptation of the materials.

\newpage

\bibliographystyle{unsrt}  
\bibliography{references}  

\end{document}